\documentclass[graybox]{svmult}

\usepackage{xparse}
\usepackage{mathptmx}       % selects Times Roman as basic font
\usepackage{mathpazo}
\usepackage{helvet}         % selects Helvetica as sans-serif font
\usepackage{courier}        % selects Courier as typewriter font
\usepackage{type1cm}        % activate if the above 3 fonts are
\usepackage{makeidx}         % allows index generation
\usepackage{graphicx}        % standard LaTeX graphics tool
\usepackage{multicol}        % used for the two-column index
\usepackage[bottom]{footmisc}% places footnotes at page bottom
\usepackage[format=plain,justification=RaggedRight,singlelinecheck=false]{caption}
\usepackage{xcolor}
\usepackage{framed}

\usepackage{pifont}% http://ctan.org/pkg/pifont
\newcommand{\cmark}{\ding{51}}%
\newcommand{\xmark}{\ding{55}}%
\usepackage{enumitem}

\usepackage{bm}
\usepackage{latexsym}
\usepackage{amsmath,amsfonts,amssymb}
\usepackage{ragged2e}
\usepackage{color}

\makeatletter
 \renewenvironment{svgraybox}
  {\begin{shaded}%
   \list{}{\leftmargin=\p@\rightmargin=\p@\topsep=\p@}%
   \expandafter\item\parindent=\svparindent
   \hskip-\listparindent}
  {\endlist\end{shaded}}
  \makeatother

\reversemarginpar

\def\eq#1{{Eq.~(\ref{#1})}}

\def\fig#1{{Fig.~\ref{#1}}}

\def\dd{\mathrm{d}}
\def\be{\begin{equation}}
\def\ee{\end{equation}}
\def\bes{\begin{eqnarray}}
\def\ees{\end{eqnarray}}
\def\f{\frac}
\def\pp{\partial}
\def\nn{\nonumber}
\def\tbf#1{{\textbf{#1}}}

\def\calO{{\mathcal O}}

\makeindex     % used for the subject index
\begin{document}

\title*{From Quantum to Classical in the Sky}

\date{\today}
\author{Suprit Singh}
\institute{Suprit Singh \at Department of Physics \& Astrophysics, University of Delhi, New Delhi 110 007 India. \\ \email{ssingh2@physics.du.ac.in}}

\maketitle 

\abstract{~~Inflation has by-far set itself as one of the prime ideas in the current cosmological models that seemingly has an answer for every observed phenomenon in cosmology. More importantly, it serves as a bridge between the early quantum fluctuations and the present-day classical structures. Although the transition from quantum to classical is still not completely understood till date, there are two assumptions made in the inflationary paradigm in this regard: \\
\\
(i) the modes (metric perturbations or fluctuations) behave classically once they are well outside the Hubble radius and, \\
\\
(ii) once they become classical they stay classical and hence can be described by standard perturbation theory after they re-enter the Hubble radius. \\
\\
We critically examine these assumptions for the tensor modes of (linear) metric perturbations in a toy three stage universe with (i) inflation, (ii) radiation-dominated and (iii) late-time accelerated phases. The quantum-to-classical transition for these modes is evident from the evolution of Wigner function in phase space and its peaking on the classical trajectory. However, a better approach to quantify the degree of classicality and study its evolution was given by Mahajan and Padmanabhan \cite{gaurang} using a \emph{classicality parameter} constructed from the parameters of the Wigner function. We study the evolution of the classicality parameter across the three phases and it turns out that the first assumption holds true, there is emergence of classicality on Hubble exit, however the latter assumption of ``once classical, always classical" seems to lie on a shaky ground.}

\newpage

\noindent If there has been a paradigm shift in cosmology, we can (with a wide consensus) say that it was the inclusion of inflation --  a period of an exponential growth at the `inception' of our universe -- even though in an ad-hoc manner 
%or what we would say in India as a \emph{jugaad}\footnote{literally meaning, \emph{a} hack.} 
to do away with the shortcomings of the hot big bang model. These were old time issues such as the flatness and horizon problems and the choice of initial conditions. However, inflation really came as \emph{deus ex machina} to explain many other significant features along with taking care of the above problems in its assorted bag of tricks. It, specifically, gave a way to unite the two scales\footnote{Usually, this means that it bridges the two scales in the high energy physics which are about $0.2$ eV and $10^{12}$ GeV or more.} which I will take as the \emph{quantum} and \emph{classical} in our universe for our interests. In inflationary cosmology -- the large scale structures, anisotropies of the cosmic microwave background, primordial magnetic fields and primordial gravitational waves all have their origins in the \emph{quantum} fluctuations that reigned during that epoch \cite{lyth}. These fluctuations are the metric perturbations on the otherwise smooth, homogenous and isotropic background, sourced by the fluctuations of an inflaton field in the simplest of models. Inflation, essentially, sets everything up in a vacuum state defined well back in the past and hence these fluctuations are \emph{quantized} and evolve very much like a quantum field on a curved background\footnote{Note that the perturbations do backreact on the geometry and also mix up on non-linear scales which makes them complicated over test quantum fields.}. That is, an infinite set of harmonic oscillator modes in Fourier space evolving in competition with the comoving Hubble length scale which decreases during the exponential expansion. Inflation, then makes the following two assumptions in this regard,

\begin{itemize}
\item[(i)] the modes behave classically once they are outside the Hubble radius and, 

\item[(ii)] once they become classical, they stay classical and hence can be described by standard perturbation theory after they re-enter the Hubble radius. 
\end{itemize}

That is, after the end of inflation, these fluctuations are taken as classical, gaussian random fields on re-entry in the radiation-dominated phase which grow due to gravitational instability. Now, the details of how the quantum-to-classical transition occurs is one of the fundamental questions that stands open in the inflationary paradigm of cosmology. Various mechanisms such as decoherence etc are invoked to explain the transition, but it is still not completely well understood. Also, without any clear quantitative measure for the classicality of the fluctuations, one has to deal with a ``two-level" description of fluctuations being either fully classical or fully quantum mechanical while the reality may lie on the middle ground. Finally, it is imperative to test the very foundations of inflation and the \emph{quantum origin} of seeds of cosmic structures. Could it be by designing a cosmological Bell-type experiment \cite{maldacena}? Or, recovering the quantum information from present-day observations of the sky \cite{martin}. Inflation, for all we know, is like modelling a black-box subcircuit using equivalent model circuits \`a la Th\'evenin and Norton, that can give similar observations\footnote{Models of inflation have parameters such as the scalar spectral index ($n_s$), amplitude ($A_s$) of scalar power spectrum, tensor-to-scalar ratio ($r$), non-gaussianity ($f_{\rm NL}$) etc. which are tested against observations.}. The actual reality could be far different from the constructed models. These questions are essential to be answered if we have to put inflation on \emph{fundamenta incocussa}.

%\begin{figure}[t!]
%\centering
%\includegraphics[scale=0.3]{blackbox}
%\caption{A Black-box subcircuit with Th\'evenin and Norton equivalents.}  
%\label{blackbox}
%\end{figure} 

The first question in the quest is, ``How to quantify the degree of classicality?". This is usually determined from the peaking of the Wigner function on the classical trajectory. However, relying on the Wigner function alone can lead to ambiguities. This is seen, for example, in Schwinger effect where there is pair production under strong external electric field. The Wigner function in this case is uncorrelated and concentrated on the classical trajectory both at very early times (when the field is in the in vacuum) and at late times (when particle number has reached an asymptotically constant limit). Thus, the peaking of Wigner function alone cannot be sufficient criterion for the emergence of classicality. For this, Mahajan and Padmanabhan \cite{gaurang} proposed using a \emph{classicality parameter} (a measure of phase space correlations) which they showed works fairly well and fits the common intuition of classicality using various examples. We employed the same construction for a test scalar field in the setting of a three stage universe \cite{suprit2013b}. The results were a bit surprising regarding the two assumptions in inflation we mentioned before and the subject of the present article is to \emph{emphasize} that the surprising results hold even for the primordial gravitational waves, i.e., the tensor modes of metric perturbations. 
\begin{figure}[t!]
\centering
\includegraphics[scale=0.6]{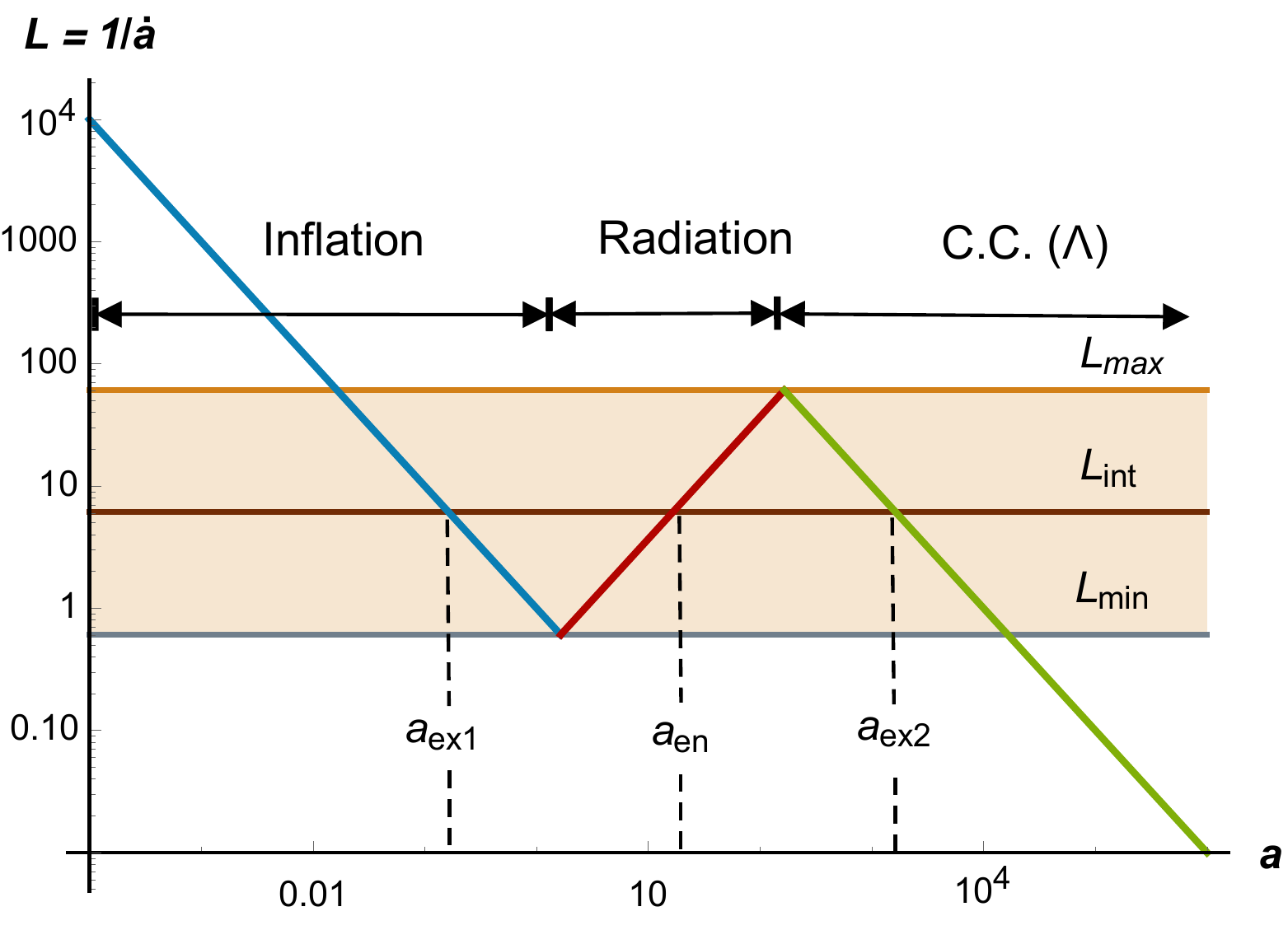}
\caption{{\bf The Cosmic Sandwich.} A (logarithmic) plot of comoving Hubble radius, $L = (da/dt)^{-1}$ where $t$ is the usual comoving time with the scale factor $a$ for $\epsilon = H_\Lambda/H_{\rm inf}= 10^{-4}$ showing the inflationary phase (decreasing), radiation-dominated phase (increasing) and in late-time cosmological constant dominated (de Sitter) phase (decreasing) where the lines have slopes: $\pm 1$. We have a natural band of two characteristic length scales, $L_{max}$ and $L_{min}$ such that any length scale within has three transition points where it goes super-Hubble, sub-Hubble and finally super-Hubble again.}
\label{comovuniv}
\end{figure} 

The idea is the following: We consider a three stage background universe\footnote{This is quite close to our real universe since matter domination lasted for a very small duration compared to the radiation-dominated stage.} that comprises of an (i) inflationary de Sitter stage, (ii) radiation-dominated stage and finally, (iii) Cosmological constant ($\Lambda$) dominated late-time accelerated stage. The transitions points maintain the continuity of scale factor and its derivative (the hubble parameter). We shall use the scale factor as the time parameter since it is monotonic.  Hence, the three stages can be expressed in terms of the comoving Hubble radius as,
%%%%%%%%%%%%%%%%%%%%%%%%%%%%%%%%%%%%%%%%%%%%%%%%%%%%%%%%%%
%%%%%%%%%%%% Inflation in a mustard seed box %%%%%%%%%%%%%%%%%%%%%%%%%%%%%%
%%%%%%%%%%%%%%%%%%%%%%%%%%%%%%%%%%%%%%%%%%%%%%%%%%%%%%%%%%
\begin{center}
\setcounter{equation}{0}  %%% HACK for getting the box equation numbers right.
\makeatletter
\let\reftagform@=\tagform@
\def\tagform@#1{\maketag@@@{(Box 1.#1\unskip\@@italiccorr)}}
\renewcommand{\eqref}[1]{\textup{\reftagform@{\ref{#1}}}}
\makeatother
\begin{figure}[t!]
\centering
\begin{svgraybox}
\begin{center}
\vspace{-.2cm}
{\bf Inflation in a \emph{Mustard} Seed}
\end{center}

\noindent In the simplest single-field models of inflation, the action is just 
\be
\label{eq1}
S = \f{1}{2} \int d^4x \sqrt{-g} \, \left[ R - (\nabla\psi)^2 - 2V(\psi)\right]
\ee
where $\psi$ is the scalar inflaton field which is minimally coupled to gravity and we have set the units in which $\hbar = c = 8\pi G_N =1$. The homogenous \emph{ansatz} is 
\be
ds^2 = -dt^2 + a^2(t) dx_idx_i\,\,; \hspace{15pt} a(t) = \exp\left(\int dt\, H(t)\right),
\ee
where we have expressed the scale factor in terms of a quasi-constant Hubble parameter, $H(t)$. The variation leads to the constraint and the dynamical equations: 
\begin{align}
3H^2 &= \rho_\psi= \f{1}{2}\dot{\psi}^2 + V,\nn\\
\dot{H} &= -\f{1}{2}\dot{\psi}^2,\nn\\
0&=\ddot{\psi} + 3 H\dot{\psi} +  dV/d\psi ;
\end{align}
of which only two are independent. A successful inflation requires the universe to expand by, at least, around 60 e-foldings for a sufficiently long time and eventually end into a radiation dominated phase. This implies that inflation is a quasi-de Sitter stage with parameters 
\begin{align}
\epsilon &= -\f{\dot{H}}{H^2} = \f{3\dot{\psi}^2}{2\rho_\psi},\hspace{5pt}
\delta = \epsilon - \f{\dot{\epsilon}}{2H\epsilon} = -\f{\ddot{\psi}}{H\psi},\hspace{5pt}
\eta = 2\epsilon -\delta
\end{align}
satisfying $\{\epsilon,|\delta|,|\eta|\}\ll1$ and $\calO(\epsilon^2,\delta^2,\epsilon\delta)\ll\epsilon$ referred to as the slow roll conditions. The dynamics at the leading order is then dictated by,
\be
3H^2 \simeq V; \hspace{5pt} 3H \dot{\psi}\simeq -  dV/d\psi
\ee 
which, given a potential, can be solved for the scale factor and the scalar field in the slow roll limit.
\end{svgraybox}
\end{figure}
\end{center}
\setcounter{equation}{0} %Hack for resetting the equation numbers. The number in the bracket points to the last equation number before box.
\vspace{-.4cm}
%%%%%%%%%%%%%%%%%%%%%%%%%%%%%%%%%%%%%%%%%%%%%%%%%%%%%%%%%%
%%%%%%%%%%%% Inflation in a mustard seed box %%%%%%%%%%%%%%%%%%%%%%%%%%%%%%
%%%%%%%%%%%%%%%%%%%%%%%%%%%%%%%%%%%%%%%%%%%%%%%%%%%%%%%%%%
\be
\label{comovlength}
   L(a) = \left\{
     \begin{array}{ll}
        (H a)^{-1} &\hspace{7pt}a \le e^{1/2} \\
        (a/ H e) &  \hspace{7pt} e^{1/2}\le a \le (e/\epsilon)^{1/2}\\
        (a\epsilon H)^{-1}  &\hspace{7pt}   a\ge (e/\epsilon)^{1/2}
     \end{array}
   \right.
\ee  
for the inflationary de Sitter ($H_{\rm inf}= H$), radiation-dominated and late-time de Sitter ($H_\Lambda = \epsilon H$) stages respectively. These are just $45^{\circ}$ lines in the logarithmic plot (see Fig.~\ref{comovuniv}). The three stages form a rich terrain and a natural sandwich of two characteristic length scales\footnote{These are actually $L(a)$ evaluated at the transition points, which themselves, are obtained using the continuity of $L(a)$ across the phases.}, $L_{max} = 1/ (H \epsilon^{1/2} e^{1/2})$ and $L_{min} = 1/(H e^{1/2})$. Then, any length scale within this band has three transition points where it goes super-Hubble, sub-Hubble and finally super-Hubble again. The modes characterized by the length scales larger than $L_{max}$ exit the Hubble radius in the first stage and remain super-Hubble after that while the modes with wavelengths smaller than $L_{min}$ remain sub-Hubble and exit the Hubble radius only in the third stage. This relative behaviour of different modes with respect to the comoving Hubble radius leads to difference in the evolution of classicality for each of them. We assume simplest model of inflation and the details are not relevant for the present analysis. But a very brief idea of background evolution in the simplest single-field, slow-roll inflationary model is given in a separate digressive box for completeness (see ref.~\cite{baumann} for the detailed physics of inflation). 

What we need are the fluctuations around the classical solution. We can decompose the metric fluctuations or perturbations on the background Friedmann solution into scalar, vector and tensor modes. These do not mix at the linear level owing to the symmetries of the background and can be studied independently of each other. We shall choose to work with the tensor modes alone in this article\footnote{The scalar mode of perturbations in a suitable gauge requires a more careful study by incorporating backreaction and the effects after re-entry in the Hubble radius \cite{suprit2016b}.}. This is for simplicity, since at linear order, the tensor perturbations are gauge-invariant and cause no backreaction to the inflationary background. Also, the tensor modes carry on without much distortion through the reheating phase and leave their imprints on the cosmic microwave background. Therefore, we begin with the quadratic action,
\be
\label{tensoractn}
S_2 = \f{1}{8} \int \dd{\bf x}\,\,\dd\tau \, \,a^2 \left[({h'}_{ij})^2 - (\nabla h_{ij})^2\right], 
\ee
which we get by expanding of the Einsten-Hilbert action (Box 1.1) to second order, using the equations of motion and restricting only to the tensor perturbations. Note that we have now shifted to using the conformal time, 
\be
\tau \equiv \int \f{dt}{a(t)}
\ee
and prime denotes derivative with respect to this time. The tensor mode action \eq{tensoractn} is same the action for a massless scalar field up to a normalization factor. Thus, we can consider the dynamics of a massless scalar field in our three stage background neglecting the two states of polarizations so that we have,
\bes
\label{actn1}
S_h &=& \f{1}{2}\int \dd^4 x\,\sqrt{-g}\,\, \pp_a h\,\, \pp^a h = \f{1}{2}\int \dd \tbf x\,\,\dd\tau\, \,a^2 \left[h'^2 - (\nabla h)^2\right].
\ees
Due to the translational invariance of the Friedmann metric, we can decompose the field into independent Fourier modes and re-express the action as 
\be
\label{reducedactn}
S_h =\f{1}{2}\int \dd \tbf k\,\int\dd\tau\, a^2 \left({h'}^2_{\tbf k} - k^2 \, h_{\tbf k}^2\right),
\ee
where $k = |\tbf k|$. We have, thus, reduced the task to tackling the dynamics of decoupled, non-relativistic harmonic oscillators with time-dependent mass and frequencies. We quantize the system in the Schr\"odinger picture. The (time-dependent) Schr\"odinger equation for the system, 
\be
i\f{\pp}{\pp\tau} \Psi(h_{\tbf k},\tau) = \left(-\f{1}{2a^2(\tau)}\f{\pp^2}{\pp h^2_{\tbf k}} +\f{1}{2}a^2(\tau)\,\,k^2\,\,h^2_{\tbf k}\right)\Psi(h_{\tbf k},\tau)
\ee
admits time-dependent, form-invariant, Gaussian states with vanishing mean given by:
\be
\label{wavefn}
\Psi(h_{\tbf k},\tau) = N \exp\left[-\alpha_k(\tau)h_{\tbf k}^2\right] = N \exp\left[-\f{a^2(\tau) k}{2}\left(\f{1- z_k}{1+z_k}\right)h_{\tbf k}^2\right].
\ee
The time evolution of the wave function is captured in the functions $\alpha_k(\tau)$ or $z_k(\tau)$ which can be seen to satisfy the equations:
\begin{align}
\label{eqnforalpha}
&{\alpha}_k' = \f{2\alpha_k^2}{a^2} - \f{1}{2} a^2 k,\\[0.5em]
{z}_k' + & \, 2\,i \,k \,z_k + \left(\f{a'}{a}\right)(z_k^2 -1) = 0
\end{align}
on substitution. These are non-linear first order Riccati-type equations and rather difficult to handle. But, we can introduce another function $\mu_k(\tau)$, defined through $\alpha_k = - (i\,a^2/2) ({\mu'}_k/\mu_k)$ which gives an equivalent easy-to-solve and familiar\footnote{In a remarkable co-incidence, the equation for $\mu_k$ is exactly the same as the field equation for $h_\tbf k$ which we would get by directly varying the action in \eq{reducedactn}.} second-order linear differential equation, 
\be
\label{eqnformu}
{\mu}_k'' + 2\left(\f{a'}{a}\right){\mu}_k' + k^2 \mu_k = 0,
\ee
The function $z_k$, referred to as the \emph{excitation} parameter, is a measure of departure from adiabatic evolution and is related to $\mu_k$ by
\be
\label{zmu}
z_k = \left(\f{k\,\mu_k +  i {\mu}_k'}{k\,\mu_k -  i {\mu}_k'}\right).
\ee
Thus, it suffices to solve for $\mu_k$ given the boundary conditions to determine the evolution of the system. Now the question is, how to \emph{quantify} the degree of classicality of such a state? We track this using a correlation function which we refer to as the \emph{classicality} parameter. It is a measure of the \emph{phase space} correlations of a system defined as
\be
\mathcal{C} \equiv \f{\langle p q\rangle}{\sqrt{\langle p^2\rangle\,\langle q^2\rangle}}.
\ee   
We can evaluate this quantity for our system using the Wigner function,
\be
\mathcal{W}(\phi_{\tbf k},\pi_{\tbf k}) = \f{1}{\pi}\exp\left[-\f{\phi^2_{\tbf k}}{\sigma_k^2} - \sigma_k^2\,(\pi_\tbf k - \mathcal{J}_k \,\phi_\tbf k)^2\right],\nn
\ee
defined in the $\phi_\tbf k - \pi_\tbf k$ phase space of the oscillator for the state in \eq{wavefn} to get,
\be
\mathcal{C}_k = \f{\mathcal{J}_k\sigma_k^2}{\sqrt{1+(\mathcal{J}_k\sigma_k^2)^2}} =  \f{2\,\mathrm{Im}(z_k)}{1 - |z_k|^2}.
\ee  
So, when the Wigner distribution is an uncorrelated product of gaussians in $\phi_\tbf k$ and $\pi_\tbf k$, i.e., $\mathcal{J}_k = 0$, which is the case for the ground state, $\mathcal{C}_k=0$ implying a pure quantum state. Otherwise, $|\mathcal{C}_k|\leq1$ with a correlated Wigner function and hence implies deviation from the quantum nature towards classical behaviour which is maximal at the extremities.

\begin{figure}[t!]
\includegraphics[width=0.47\textwidth]{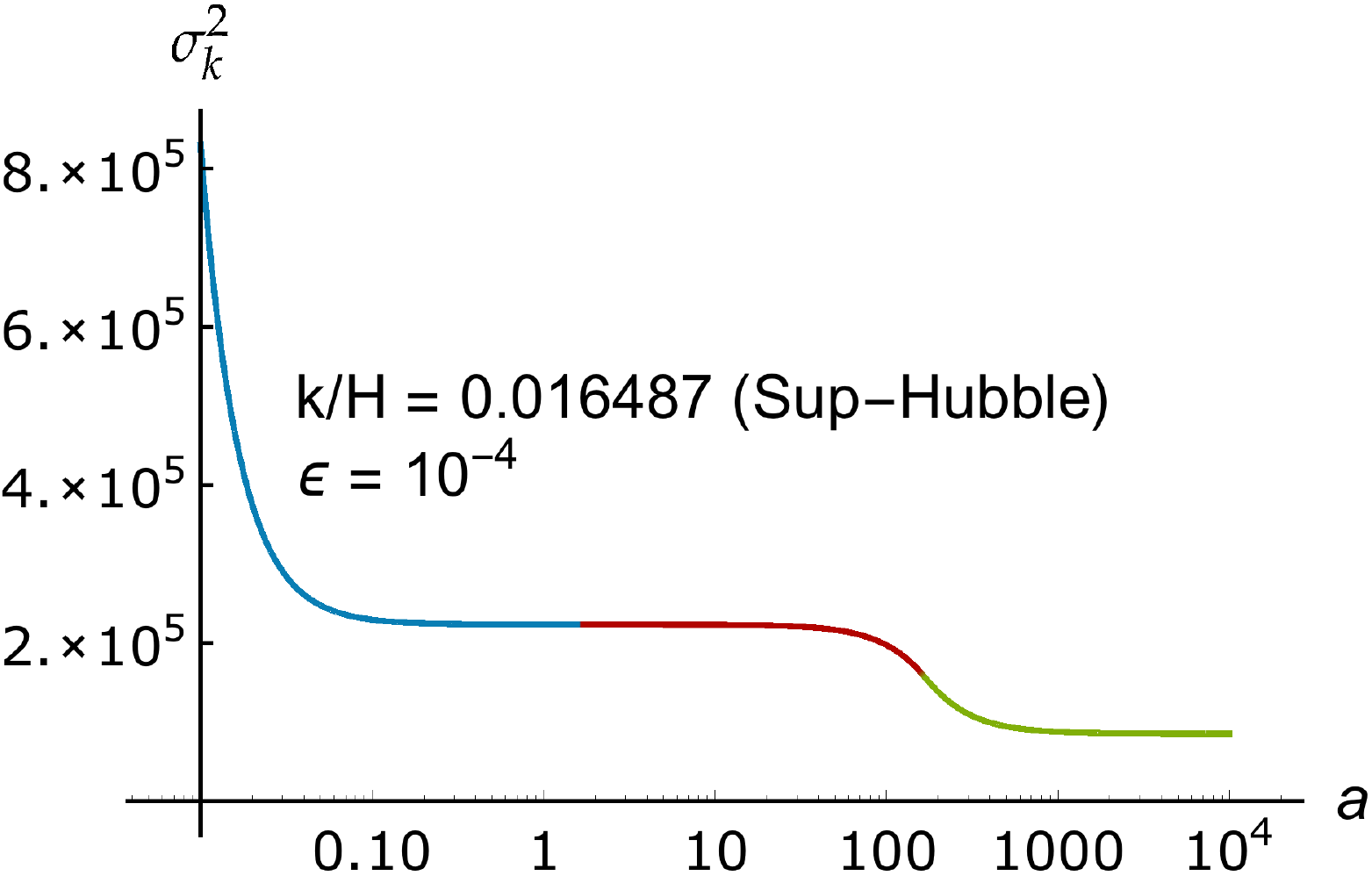}
\includegraphics[width=0.47\textwidth]{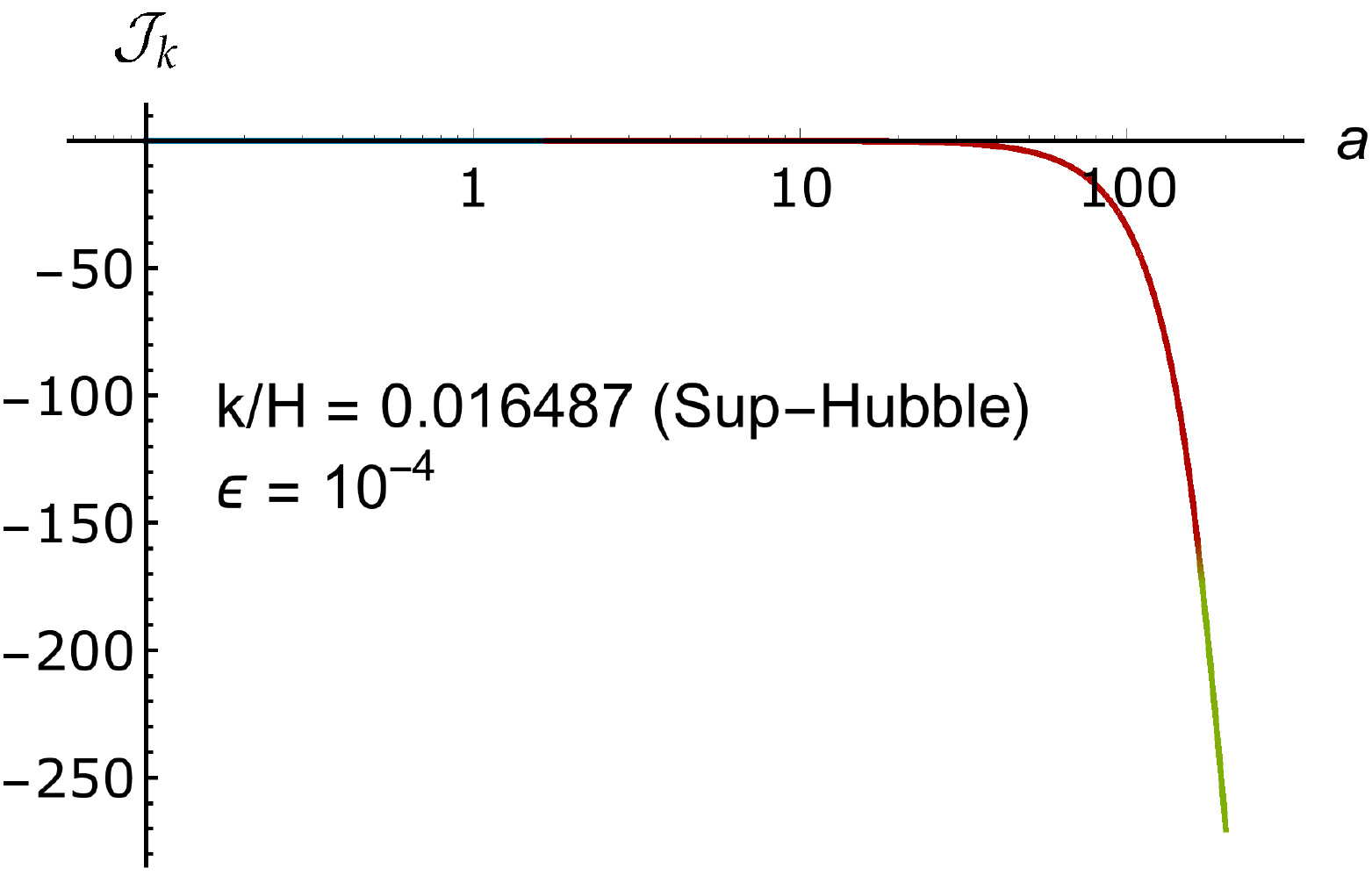}\\
\includegraphics[width=0.97\textwidth]{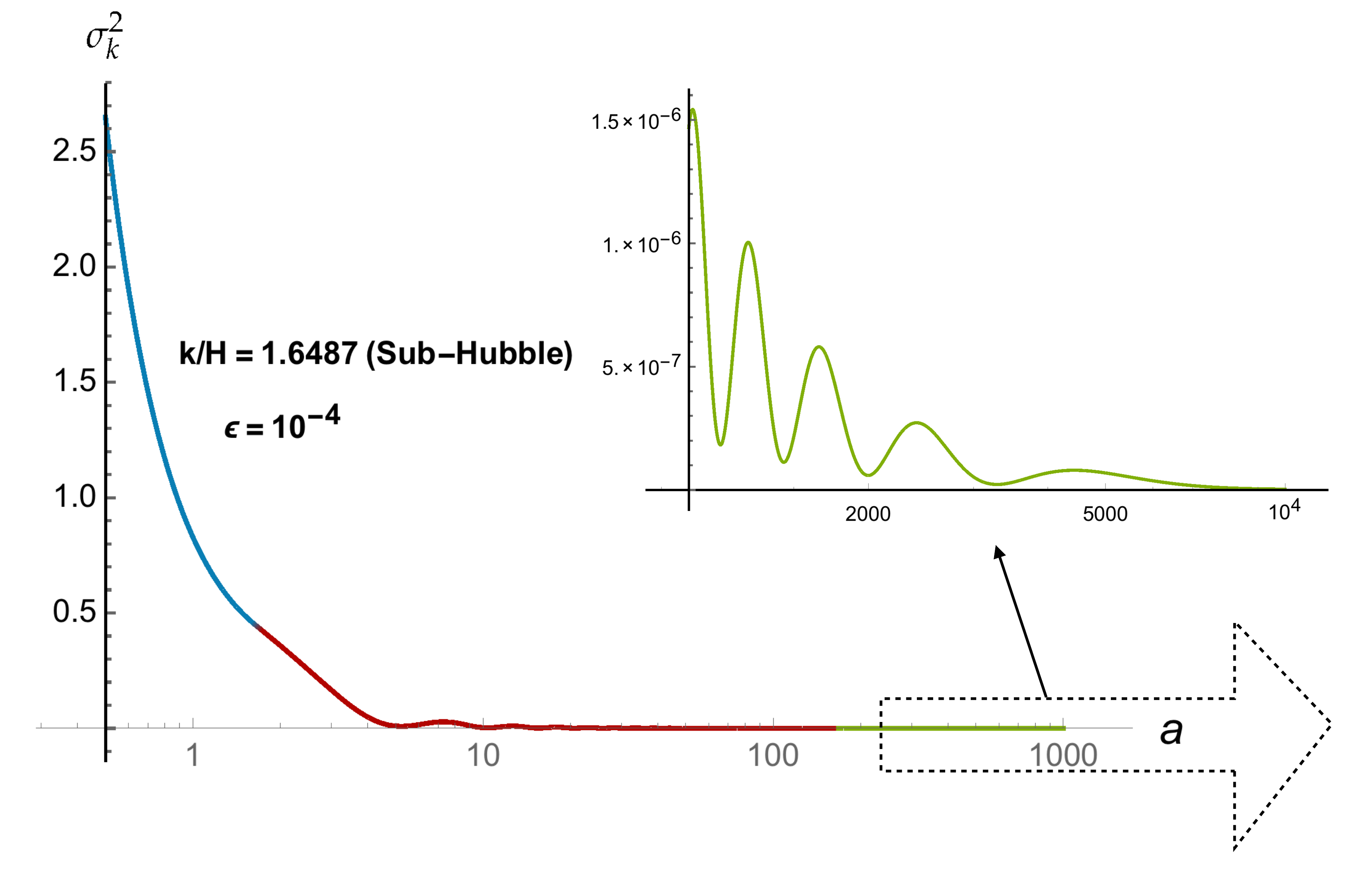}
\caption{Evolution of the parameters $\sigma^2_k$ and ${\cal J}_{k}$ of the Wigner function with the scale factor for $\epsilon = 0.0001$ for super- and sub-Hubble modes. The color scheme is as follows: early inflationary phase (blue), radiation dominated phase (red) and late-time de Sitter phase (green).}  
\label{sigmascri}
\end{figure}
\begin{figure}[t!]
\includegraphics[width=\textwidth]{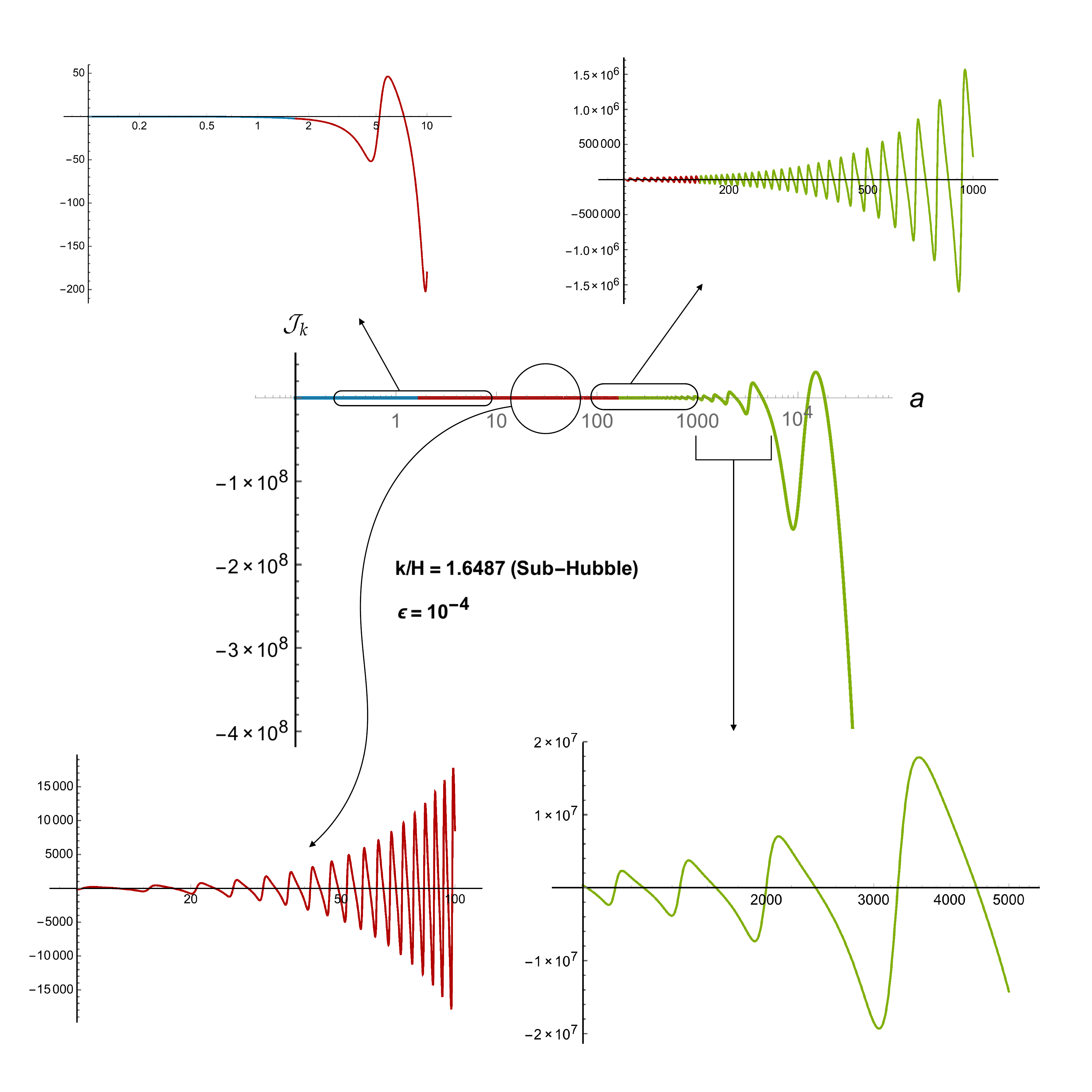}
\caption{Evolution of the parameter ${\cal J}_{k}$ of the Wigner function with the scale factor for $\epsilon = 0.0001$ for the sub-Hubble mode along with zoomed-in plots showing oscillations at different scales.}  
\label{sigmascri1}
\end{figure}

\begin{figure}[p]
\centering
\vspace{10pt}
\includegraphics[scale=0.25]{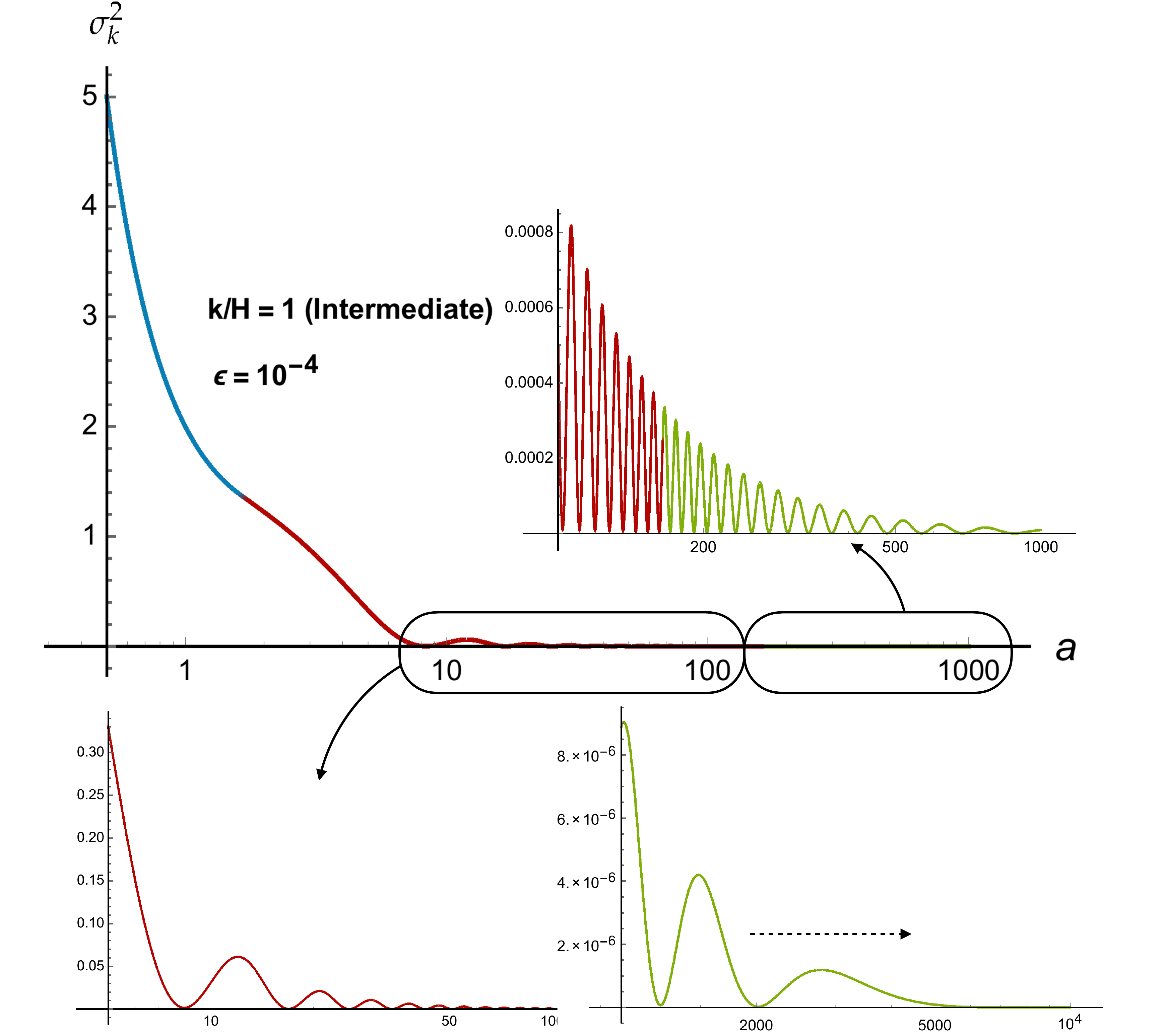}\\
\vspace{5pt}
\includegraphics[scale=0.25]{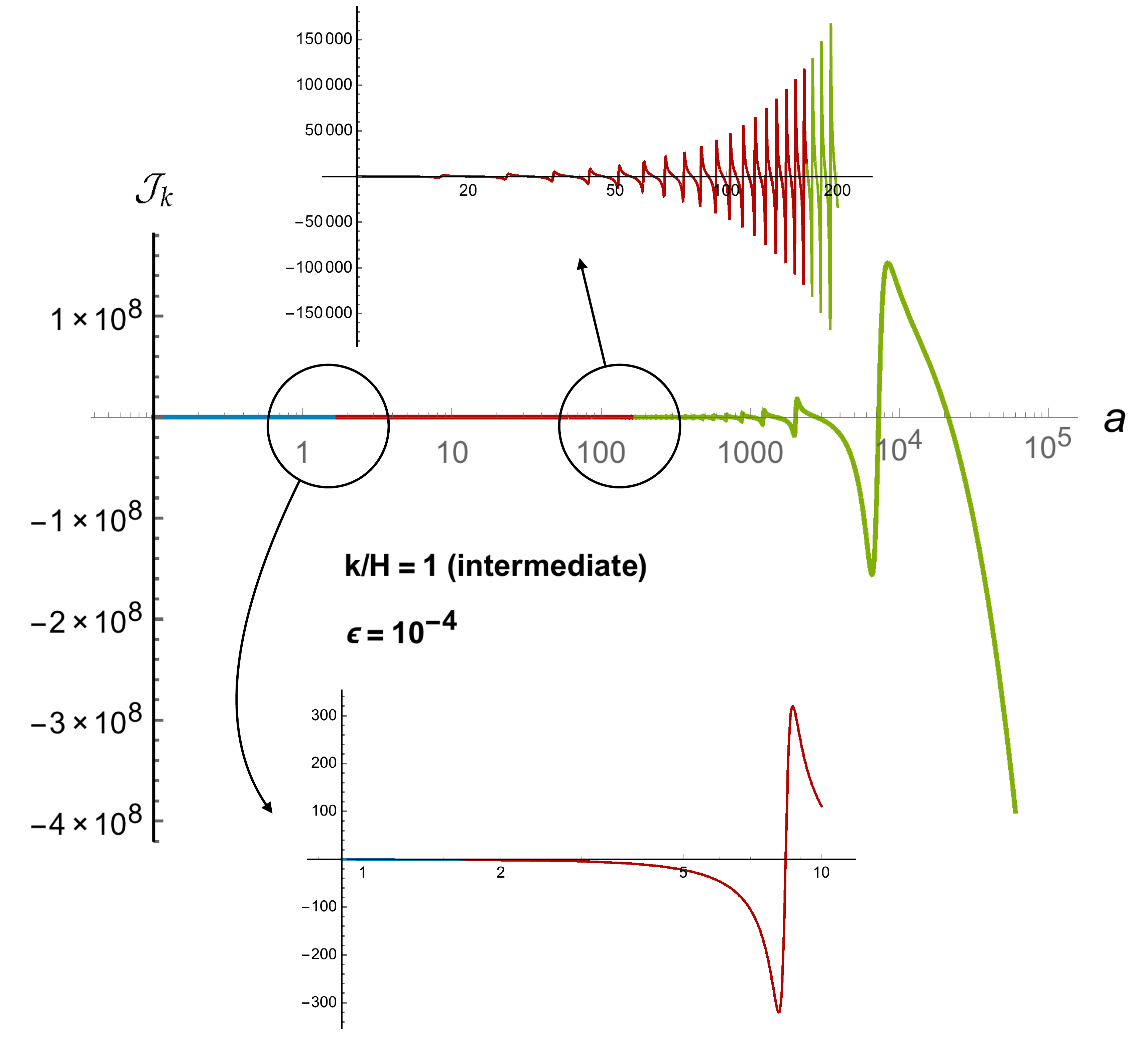}
\caption{Evolution of the parameters $\sigma^2_k$ (above) and ${\cal J}_{k}$ (below) of the Wigner function with the scale factor for an intermediate mode and $\epsilon = 10^{-4}$ along with zoomed-in plots at different times.}  
\label{sigmascri2}
\end{figure}

\begin{figure}[t!]
\includegraphics[width=0.333\textwidth]{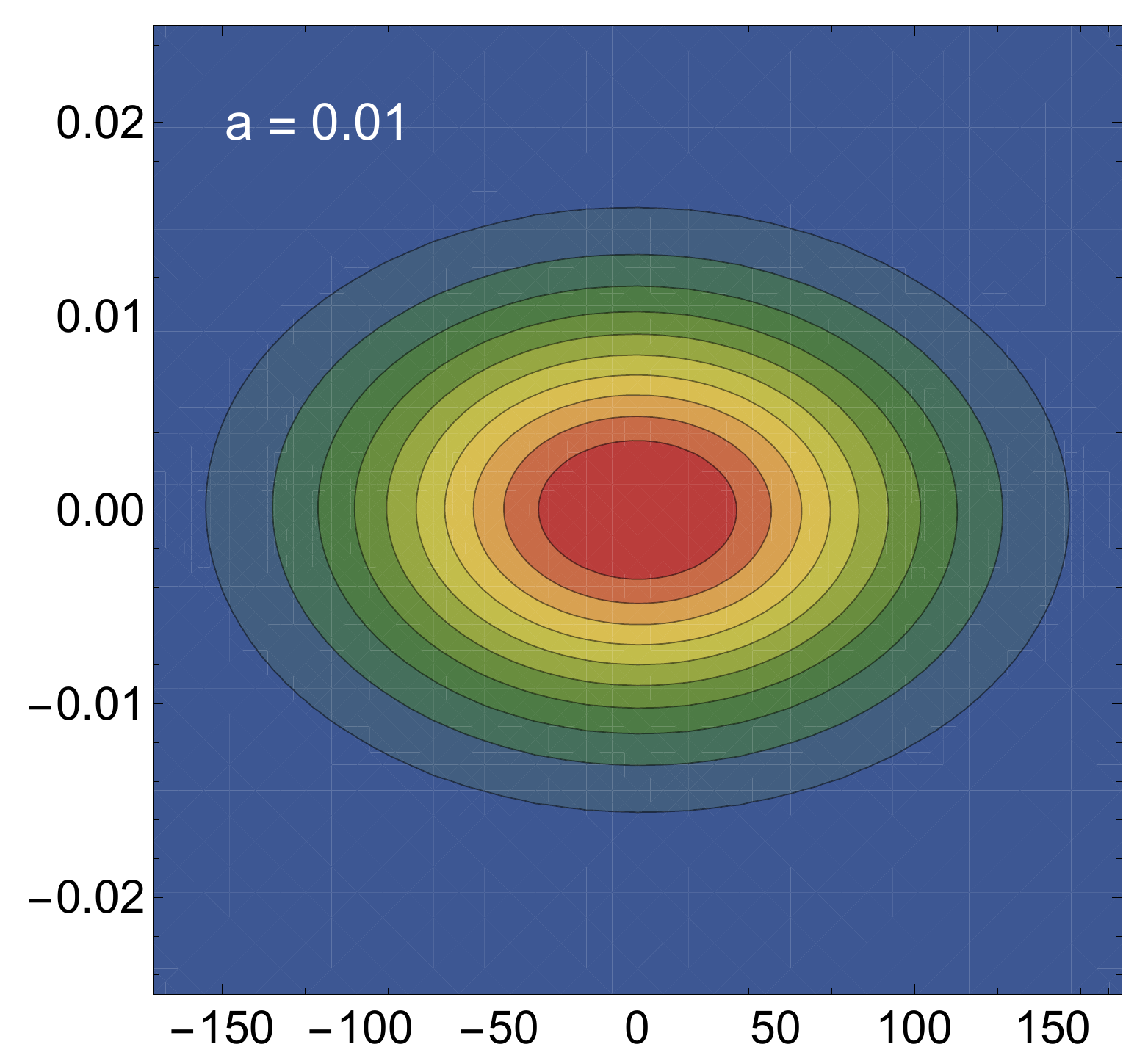}\hfill
\includegraphics[width=0.333\textwidth]{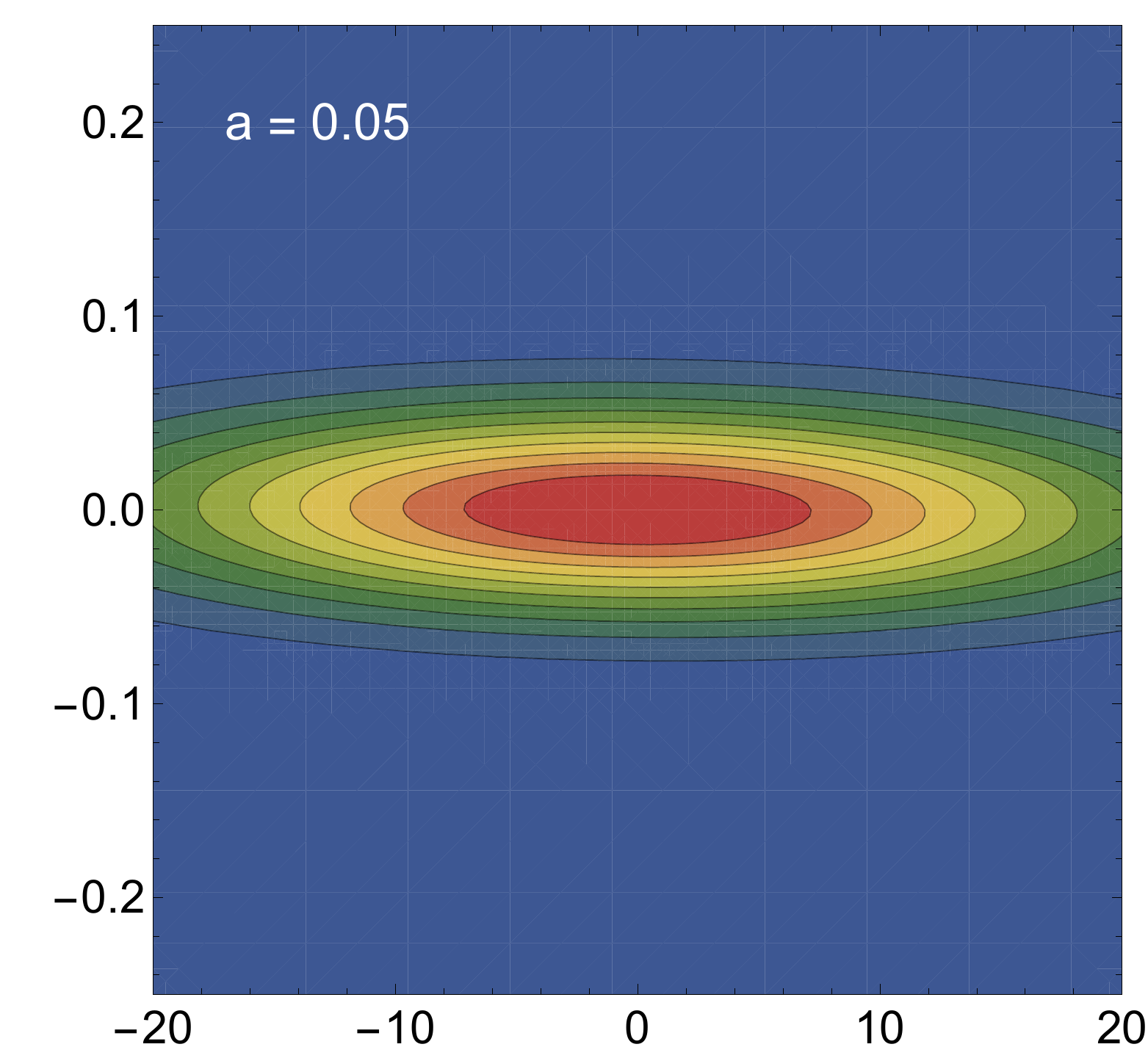}\hfill
\includegraphics[width=0.333\textwidth]{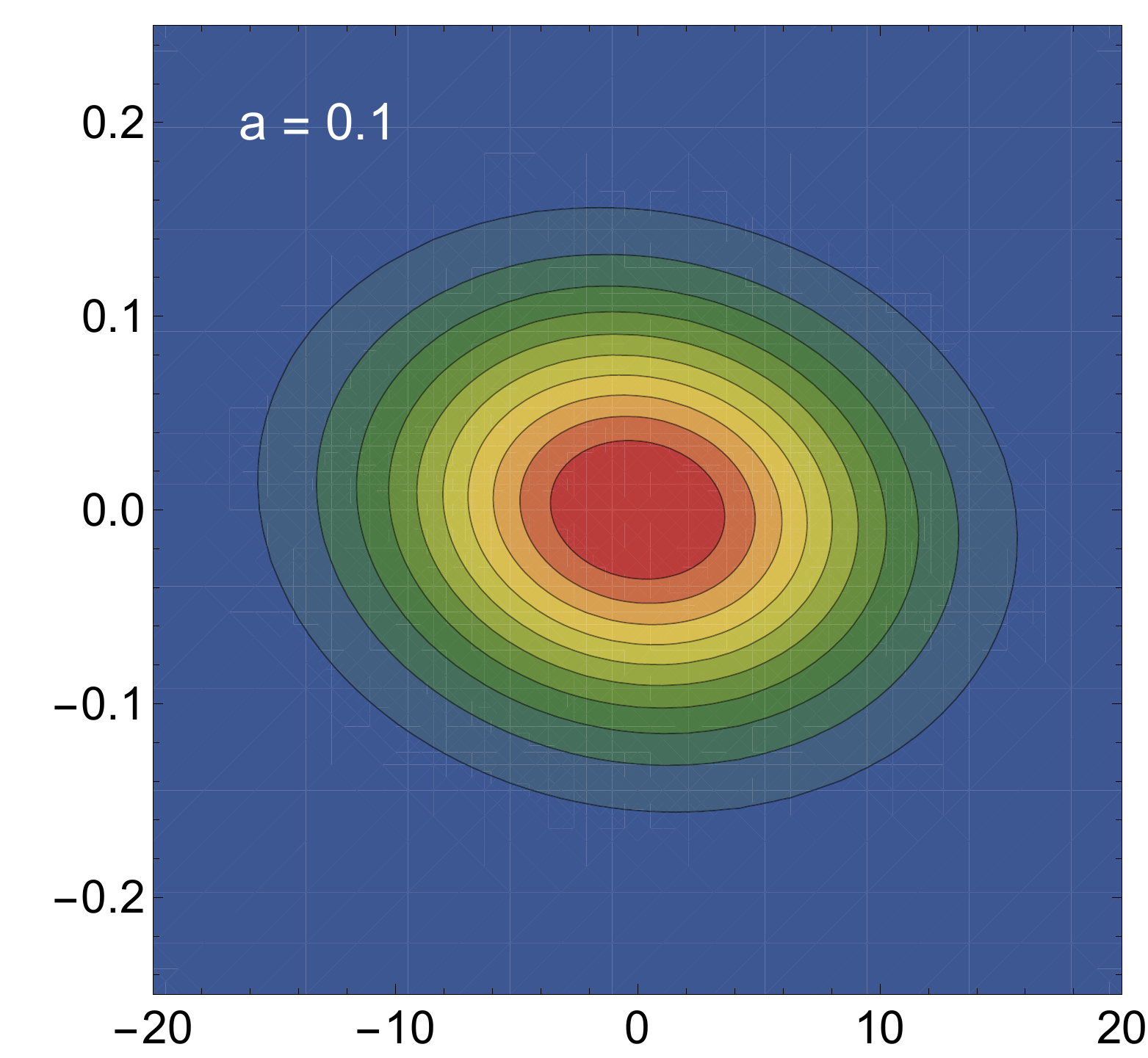}
\includegraphics[width=0.333\textwidth]{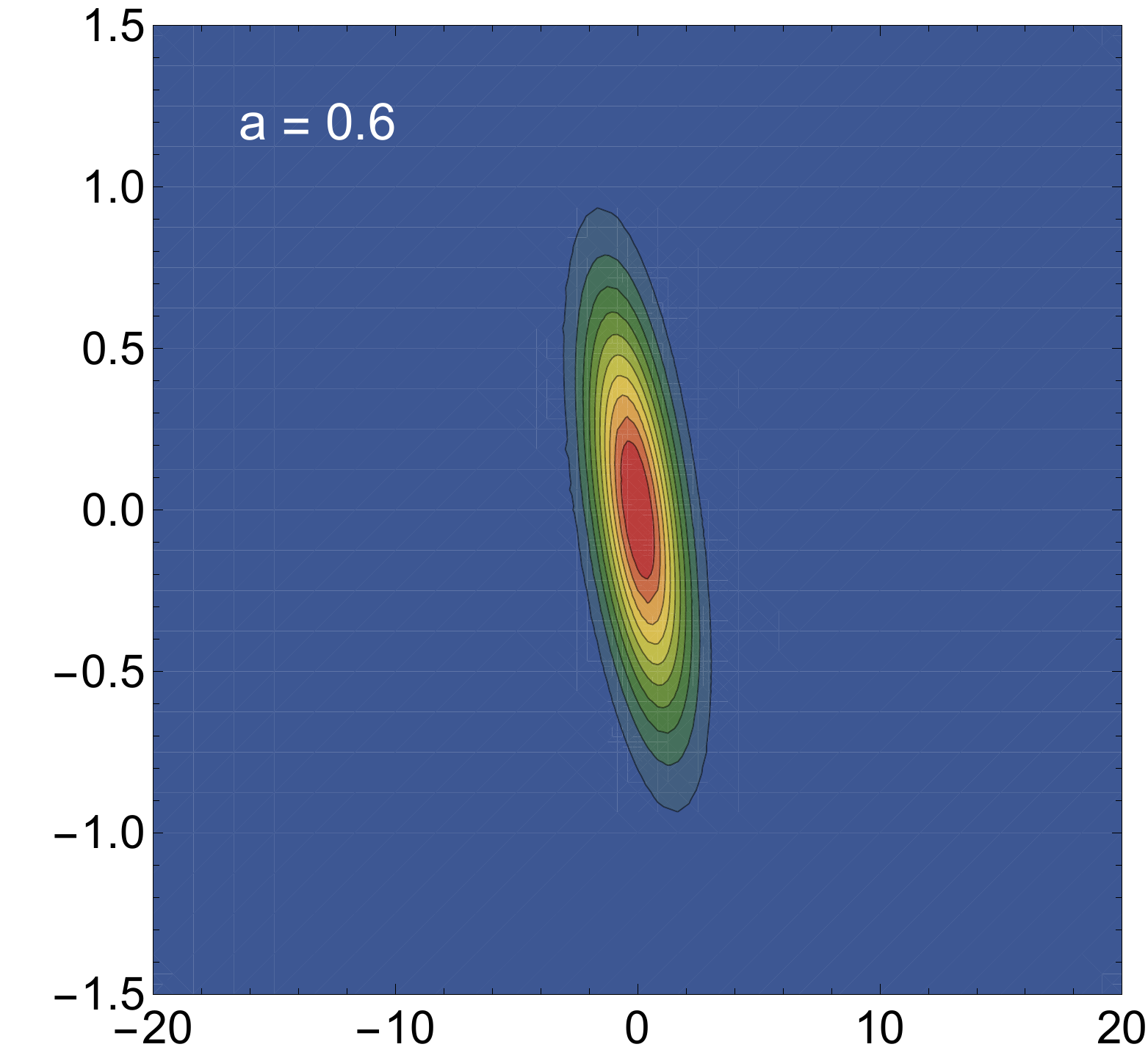}\hfill
\includegraphics[width=0.333\textwidth]{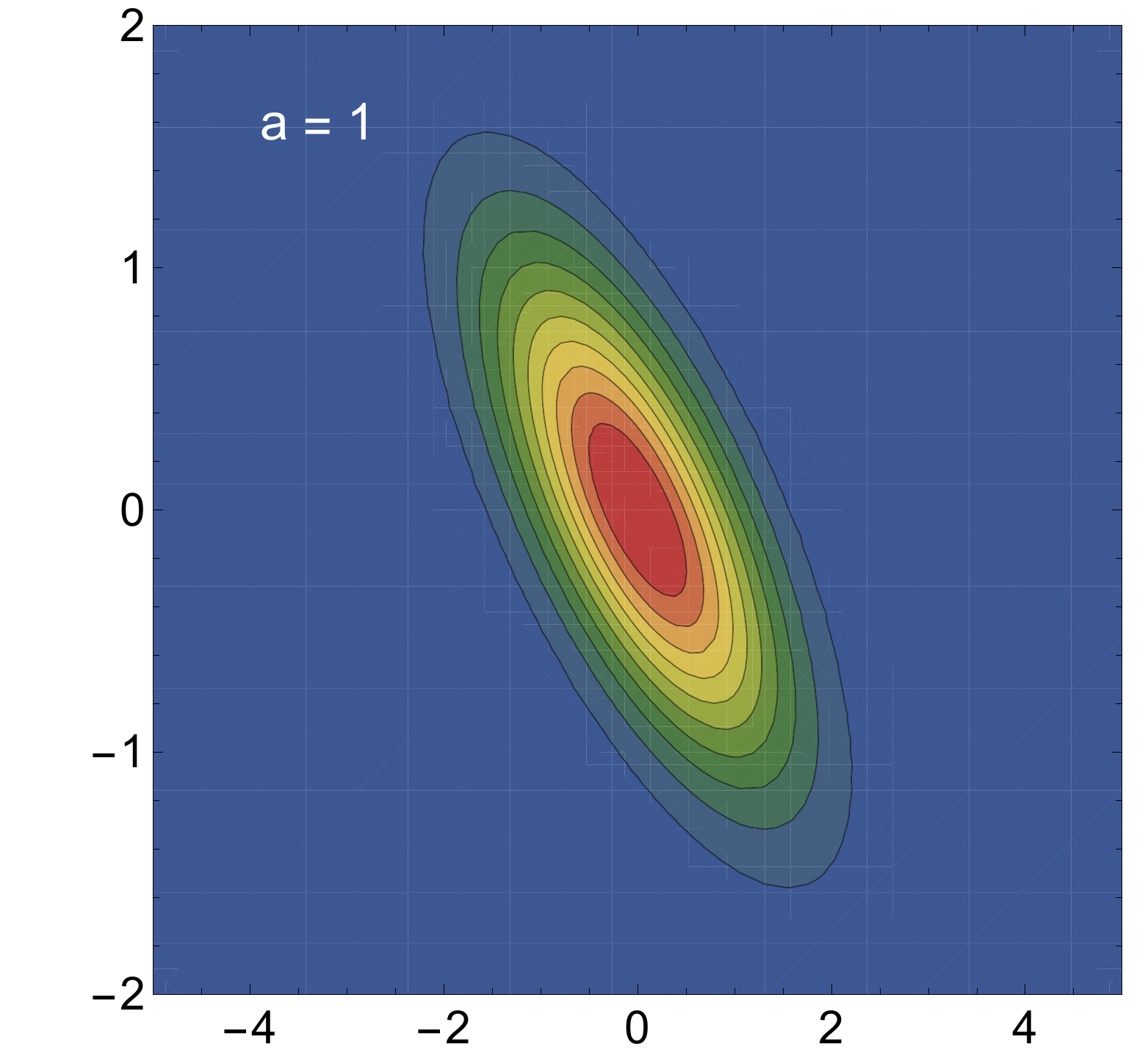}\hfill
\includegraphics[width=0.333\textwidth]{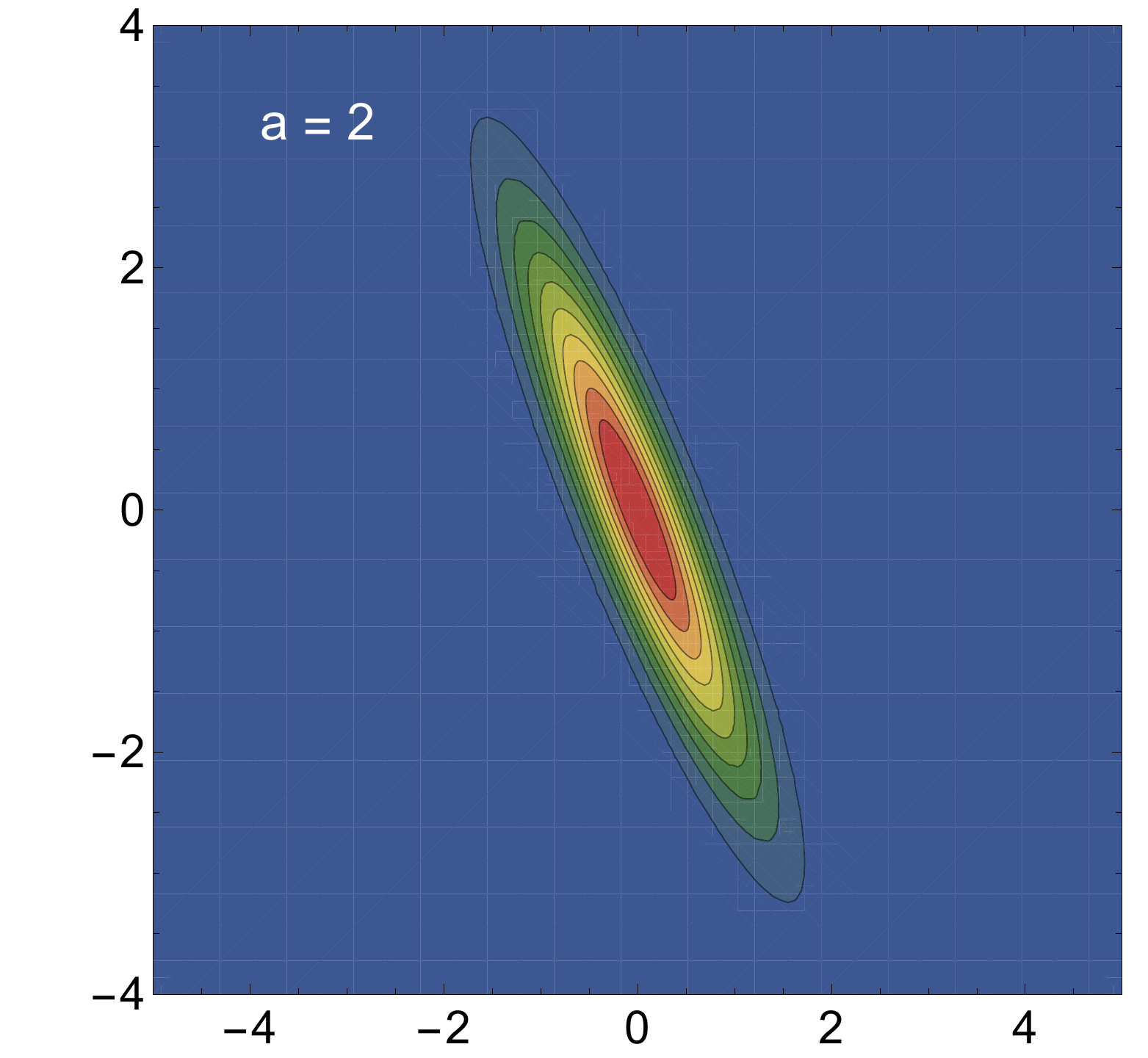}
\includegraphics[width=0.333\textwidth]{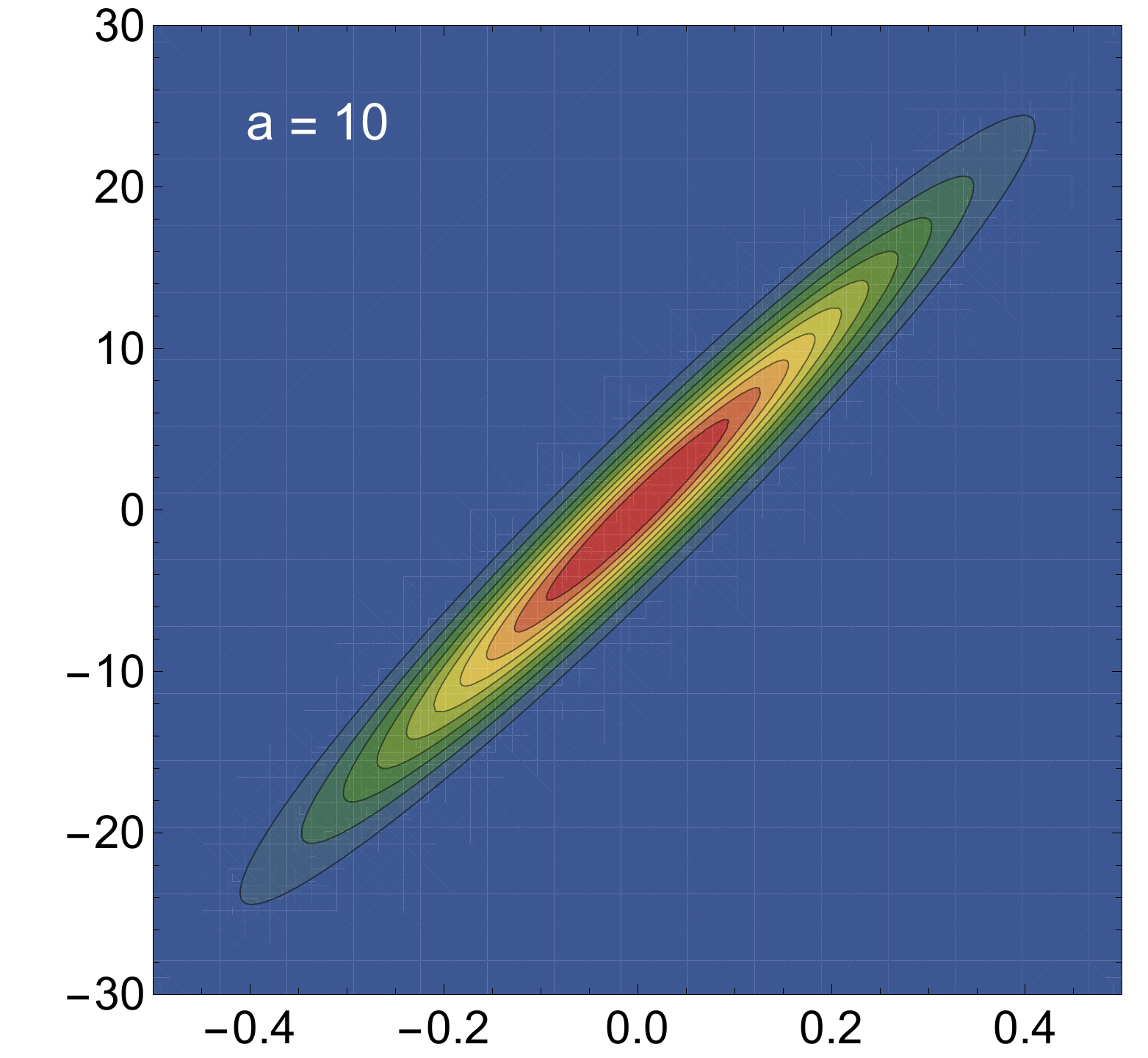}\hfill
\includegraphics[width=0.333\textwidth]{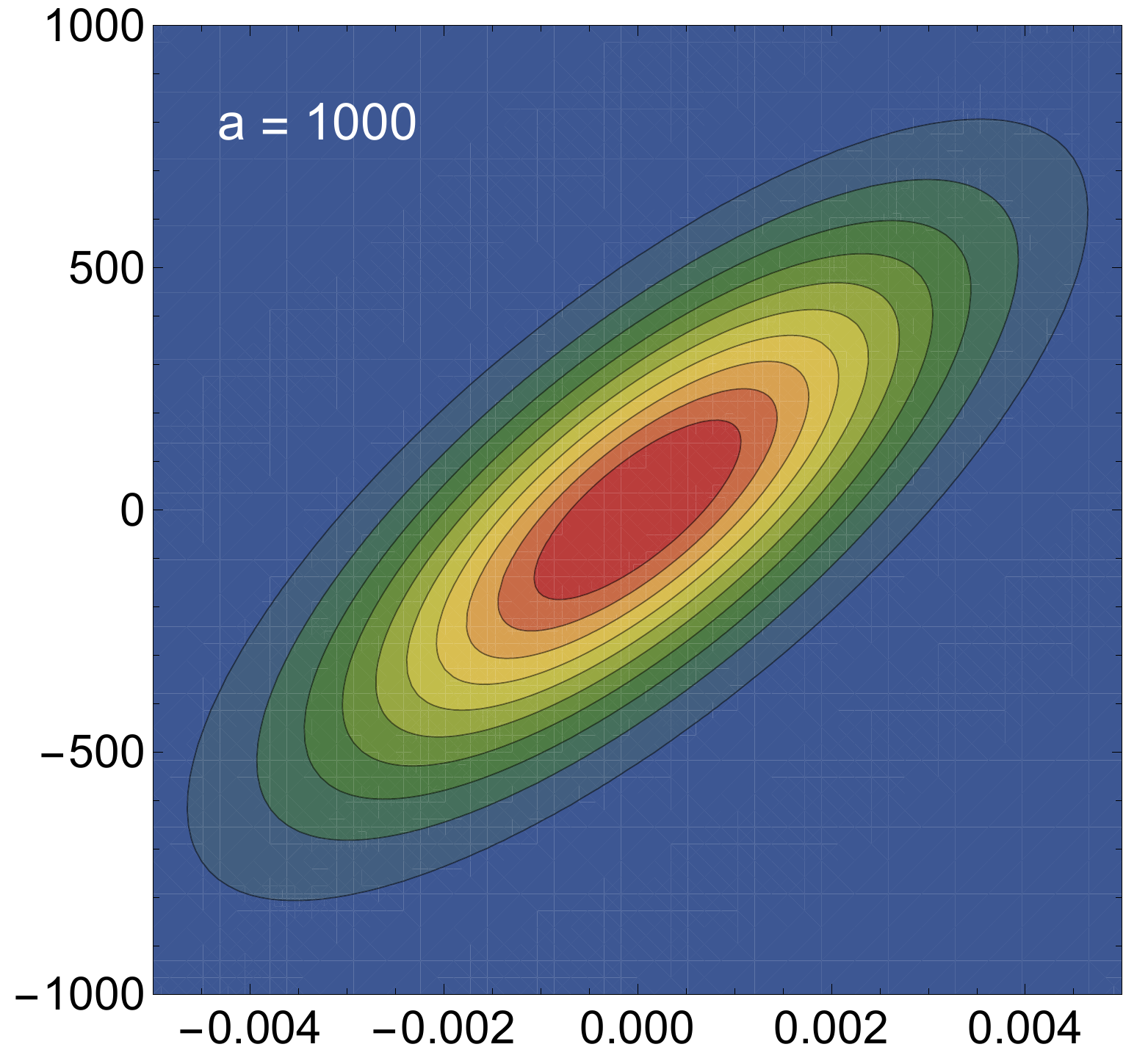}\hfill
\includegraphics[width=0.333\textwidth]{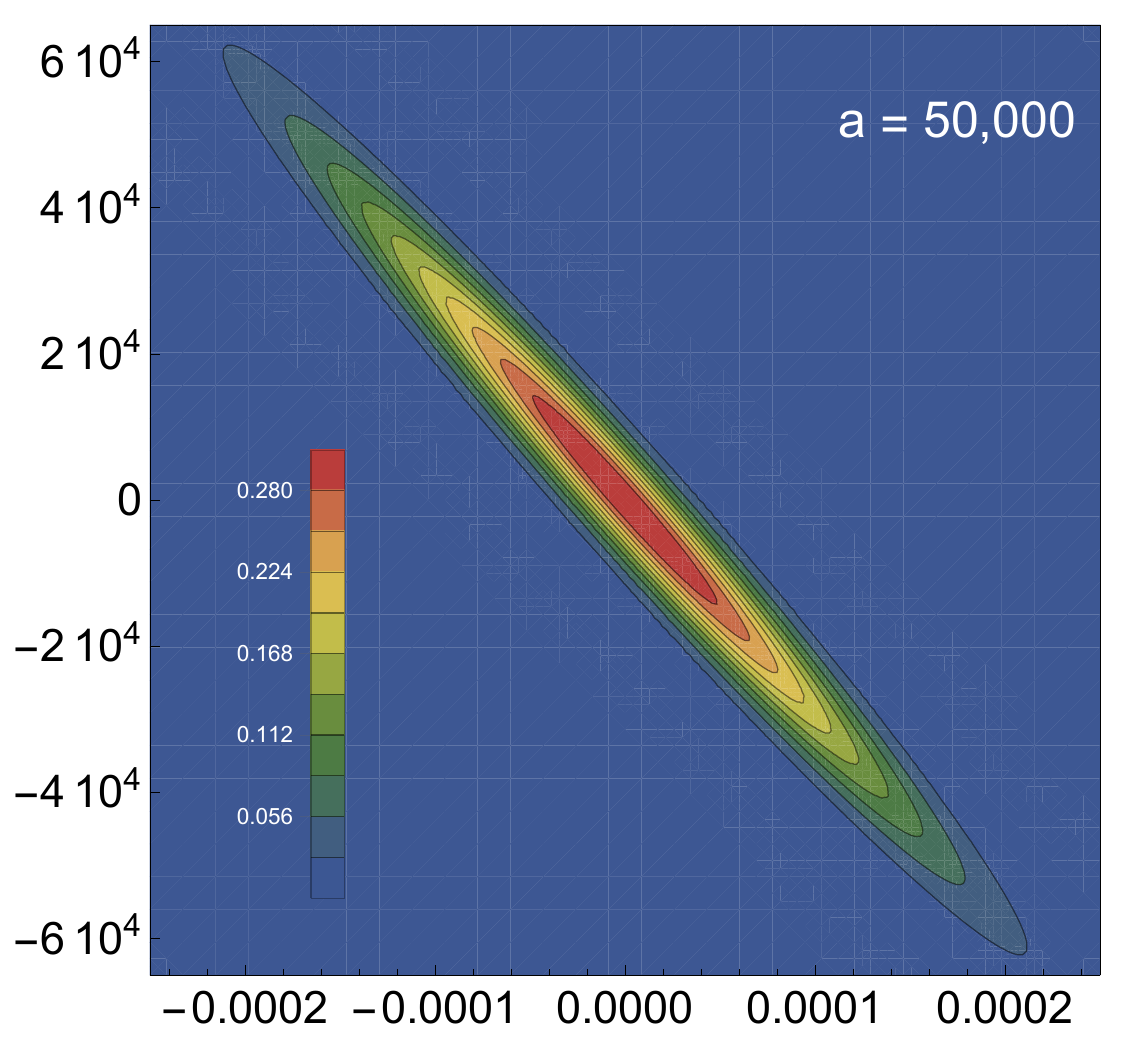}
\caption{Snapshots depicting the evolution of the Wigner function in the $\phi_k-\pi_k$ phase space for the intermediate mode ($k/H=1$) and $\epsilon = 0.0001$.}  
\label{wign}
\end{figure}

All that is required now is to solve \eq{eqnformu} for $\mu_k$ with appropriate initial conditions. We choose the standard Bunch-Davies vacuum initial condition during infation and the mode functions in the subsequent stages are stitched together by demanding the continuity of $\mu_k$ and its derivative. All other quantities can be obtained once $\mu_k$ is known throughout the history of the universe. It is to be noted that, in reality, we have $\epsilon = \mathcal{O}\left(10^{-53}\right)$ which is a very, very small number and the only feasible study is through appropriate approximations in the analytical computation (see ref. \cite{suprit2013b} for details). Here we present only the numerically computed results taking $\epsilon = 0.0001$ for visual and conceptual clarity. We show the evolution of parameters of the Wigner function in Figs.~\ref{sigmascri}, \ref{sigmascri1} and \ref{sigmascri2} and the snapshots depicting the evolution of Wigner function in phase space in \fig{wign}. 

It is evident from the plots of the functions $\sigma^2_k$ and $\mathcal{J}_k$, the Wigner function starts uncorrelated with  $\sigma^2_k \rightarrow \infty$ and $\mathcal{J}_k \rightarrow 0$ at early times for all the modes when they are at sub-Hubble scales. The Wigner function is peaked highly on the $\phi_k$-axis (the first plot in \fig{wign}). Further on $\sigma_k$ decreases sharply and $\mathcal{J}_k$ increases to negative values implying increasing anti-correlation. The Wigner function now starts to spreads out in the direction of momentum axis and lessens its spread on the position axes. The dynamics after this is strictly governed by the competition of the mode with the Hubble radius. The modes that exit the Hubble radius and go super-Hubble see a saturation of $\sigma^2_k$ and a monotonic increase of $\mathcal{J}_k$ to negative infinity. The Wigner function peaks on the momentum axis implying classical behaviour. For the modes which remain sub-Hubble until the final third stage, the limits $\sigma^2_k\rightarrow 0$ and $\mathcal{J}_k\rightarrow -\infty$ are reached non-trivially with oscillations in between during the radiation-dominated epoch. 

The most interesting to us are the intermediate modes that exit the Hubble radius during inflation and then re-enter in the radiation-dominated epoch. The plots in \fig{sigmascri2} show a slight saddle of saturation as the mode exits the Hubble radius but then on re-entry the oscillations set in for both $\sigma_k$ and $\mathcal{J}_k$ which damp to zero and increase to negative infinity respectively. The corresponding evolution of Wigner function in the phase space is also highly non-trivial. On towards Hubble-exit, the plots for $a=0.05, \,0.1,\, 0.6$ and $1$ shows the swift peaking of Wigner function on momentum axes which persists for $a=2$ when the mode is super-Hubble even though the Universe has made a transition to the second stage. However after re-entry, due to the oscillations, the Wigner function even though peaked on momentum axis, keeps on twisting and turning until the exit in the third stage to finally settle in the configuration shown for $a=50,000$. 
But clearly, the notion and quantifying the degree of \emph{classicality} is not so easy relying on the Wigner function alone. Its parameters lie in large ranges, for example, $\sigma^2_k$ is very large at early-times and $\mathcal{J}_k$ goes to negative infinity at late-times with oscillatory behaviour in between, leading to a very complicated evolution of the Wigner function in the phase space. This is where the classicality parameter helps being restricted to a finite domain [$-1,1$] by construction and as shown in \fig{classicality} for different $k/H$ values. We see the following evident features: 
\begin{figure}[t!]
\includegraphics[width=0.5\textwidth]{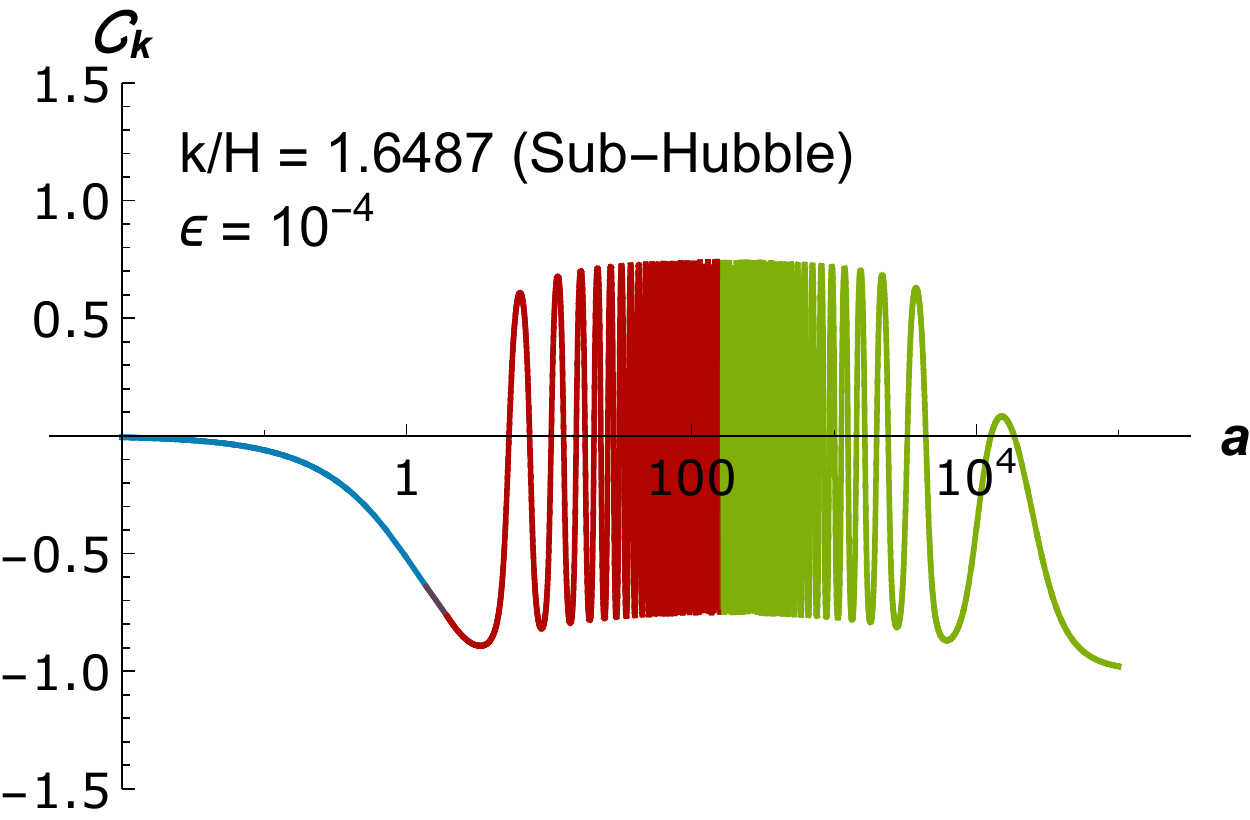}\hfill
\includegraphics[width=0.5\textwidth]{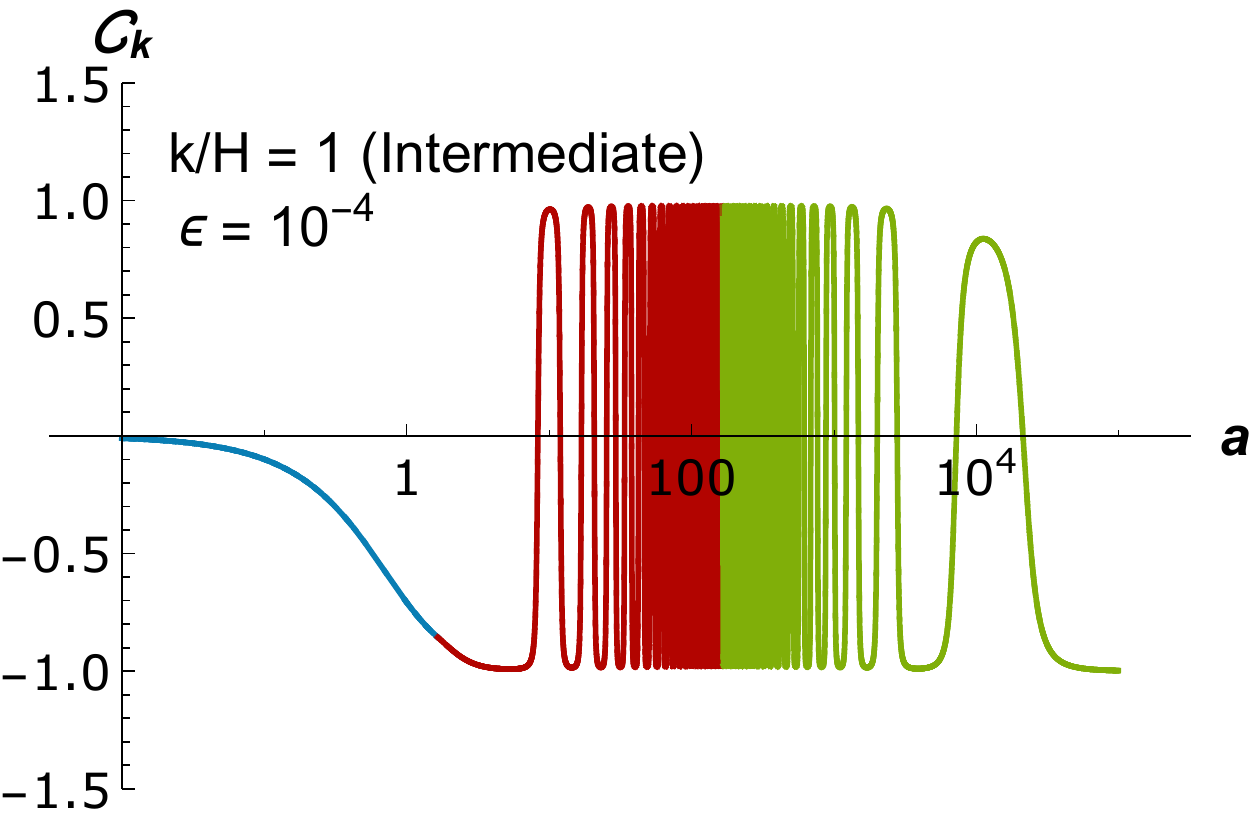}
\centering \includegraphics[width=0.5\textwidth]{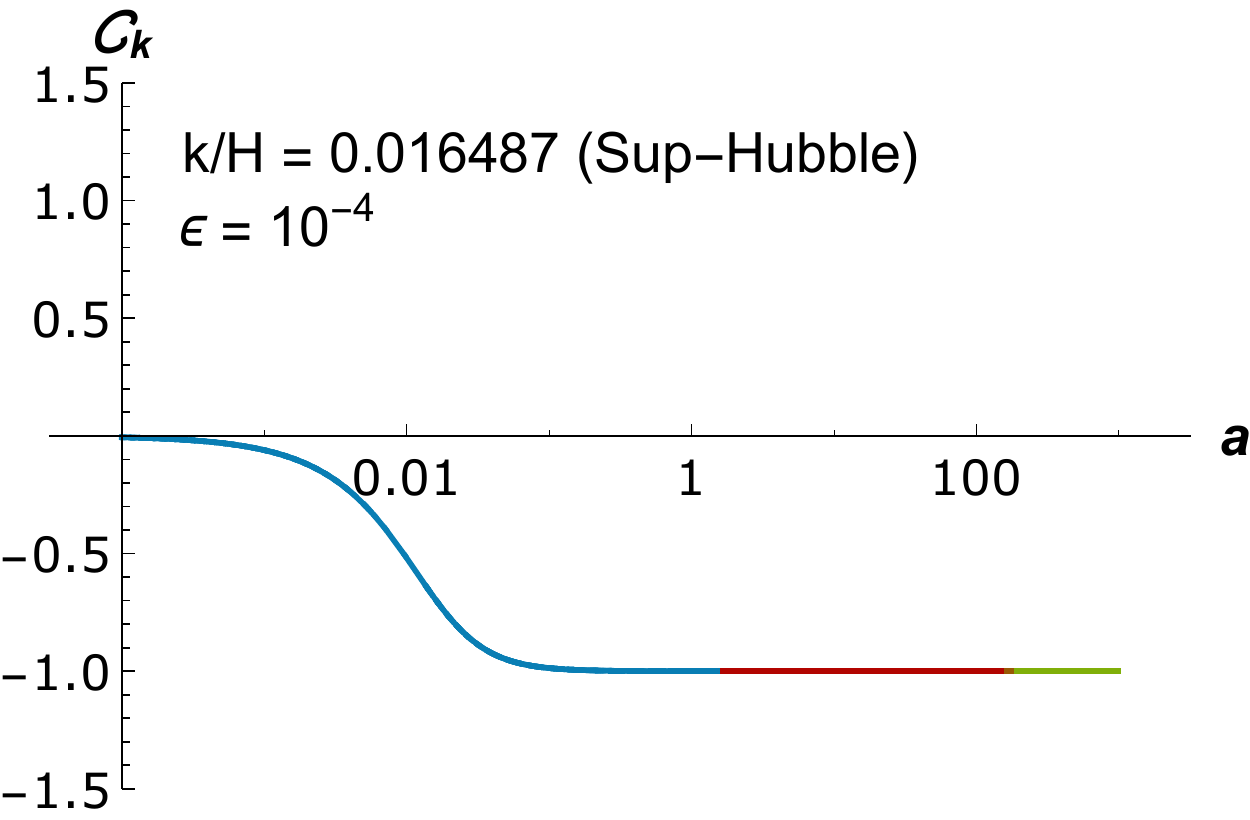}\hfill
\includegraphics[width=0.5\textwidth]{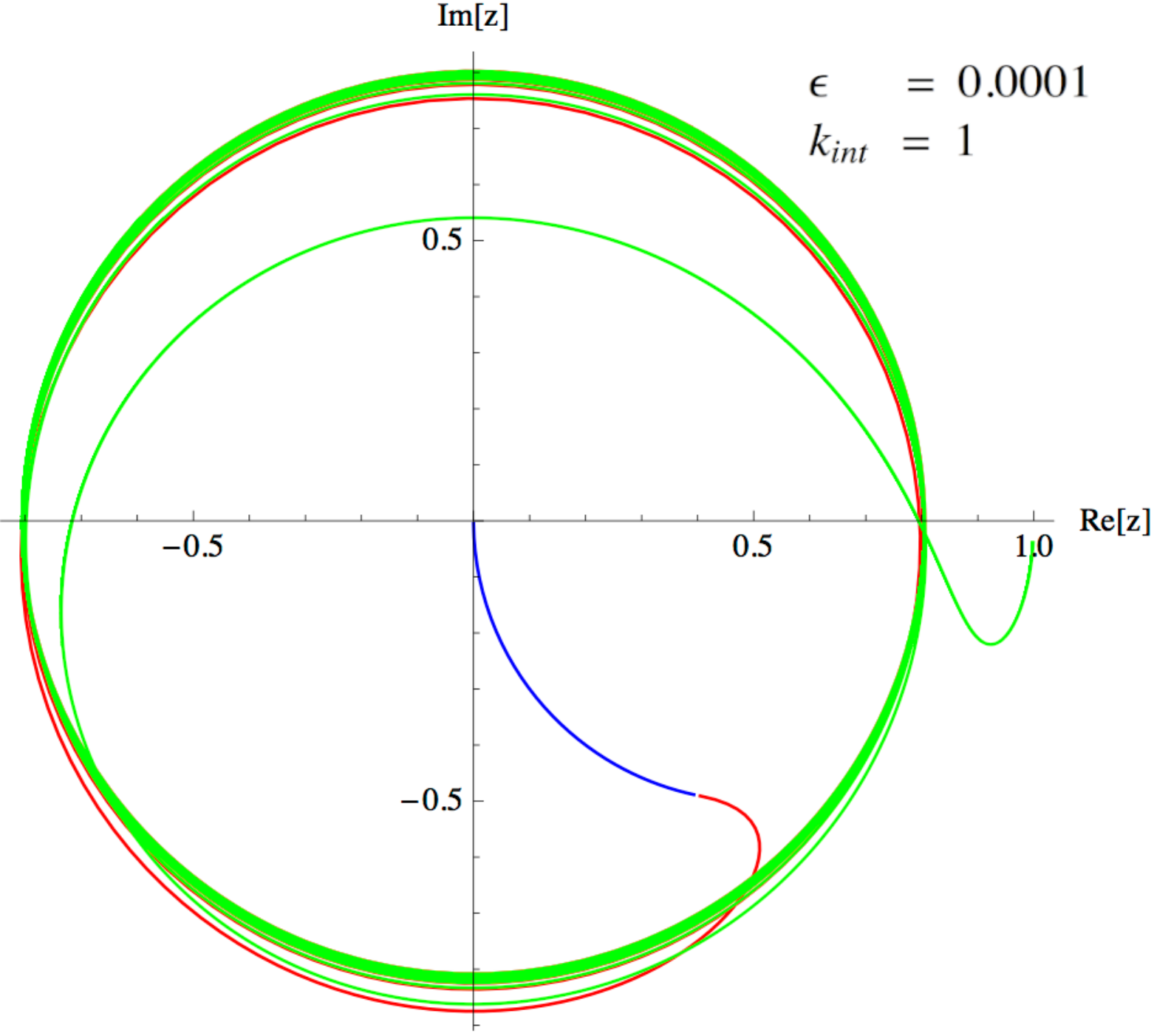}
\caption{Evolution of the classicality parameter ${\cal C}_{k}$ and the $z_k$ with the scale factor for $\epsilon = 0.0001$. We clearly see that $|{\cal C}_{k}|\rightarrow 1$ whenever a mode exits the Hubble radius, however on re-entry, there are significant oscillations which seem to imply the dynamics is away from a completely classical description. The evolution is clearly non-adiabatic evident from non-zero values of $z_k$.}  
\label{classicality}
\end{figure}
\begin{itemize}

\item The classicality parameter ${\cal C}_k$ starts from zero in the beginning of inflationary phase (indicating quantum origin) and after the Hubble exit whether it is in the early or late-time de Sitter stage, ${\cal C}_k \rightarrow -1$ and modes behave classically. 

\item For any mode with $k = k_{int}$ which lies within the $[k_{min},\,k_{max}]$ band, we have a classical description near the end of the inflationary phase as ${\cal C}_k \rightarrow -1$ but as the universe makes a transition to radiation phase, it starts oscillating. These oscillations last all through the radiation phase and the late-time de Sitter phase till the mode exits the Hubble radius and then one finds  ${\cal C}_k$ saturates at $-1$. Thus in between, when the mode is sub-Hubble, the oscillations imply that the system is away from a \emph{completely} classical description.

\item The sub-Hubble mode does not reach $-1$ in the first two phases but once it exits the Hubble radius in the late-time de Sitter phase it reaches that value. On the other hand, the super-Hubble mode, once it exits from the initial de Sitter phase always remains super-Hubble and saturates with ${\cal C}_k \rightarrow -1$. 

\item This non-trivial behaviour of classicality parameter has a reason which is evident from the plot of $z_k$ in \fig{classicality}. The function $z_k$ is a complex quantity and its non-zero value measures departure from the adiabatic evolution of modes with respect to the background. There is a delay in the change of course of $z_k$ and the background evolution. The system persists in the previous dynamics even though the Universe has made a transition to the next stage. Further, the rotor-like behaviour of $z_k$ is what causes oscillations in the Classicality parameter. The rotations start in the radiation-dominated phase and persist in the late-time phase until the mode exits the Hubble radius.    

\end{itemize}

\noindent Conclusively then, we have, for the two aforementioned \emph{assumptions} of inflation, 
\begin{enumerate} 
\item[\cmark] the emergence of classicality on Hubble exit checked,
\item[\xmark] but ``once classical, always classical" does not seem to hold.
\end{enumerate}

\noindent What does it mean for the present-day observations that assume a classical-only description of the modes that re-enter the Hubble radius? Certainly, we require a re-think and re-check of the standard procedures. Even possible effects on the non-gaussianity parameter $f_{\rm NL}$ because of the non-classical behavior of perturbations after re entry. However, it doesn't stop here and brings more questions to the table such as: How will the above conclusion fare with the back-reaction included in and for the scalar perturbations? How does it compare with other constructs that describe the quantum-to-classical transition such as the \emph{quantum discord} \cite{discord}? What does it mean to have a quantum nature intertwined with the classical and can this quantum nature have any imprints in the cosmic microwave background which may be extracted \cite{lim}?

As much as it is important to test inflation to its roots, it is a poetic quest to truly understand our \emph{quantum origins} if it indeed turned from quantum to classical in the sky!\\

\footnotesize
\noindent The research of S.S. is supported by Dr. D. S. Kothari Postdoctoral fellowship from University Grants Commission, Govt. of India. I also thank Sujoy and Paddy for the collaboration leading to ref. \cite{suprit2013b}.

\end{document}